# Differentiation method for localization of Compton edge in organic scintillation detectors

M.J. Safari*[1], F. Abbasi Davani[2], H. Afarideh[1]

[1]*Department of Energy Engineering and Physics, Amir Kabir University of Technology, PO Box 15875-4413, Tehran, Iran*
[2]*Radiation Application Department, Shahid Beheshti University, PO Box 1983963113, Tehran, Iran*
*email: mjsafari@aut.ac.ir

**Abstract**. This paper, presents a simple method for accurate calibration of organic scintillation detectors. The method is based on the fact that differentiating the response function leads to accurate estimation of the Compton edge. The differentiation method in addition to the location of the Compton edge, gives insights into the parameters of the folded Gaussian function which is useful for determination of the energy resolution. Moreover, it is observed that the uncorrelated noise in the measurement of the response function does not impose significant uncertainties in the evaluations. By simulation of the bounded electrons and considering the Doppler effects, we are able to calculate a first estimation for the intrinsic Doppler resolution of a plastic scintillator, benefiting from the capability of the differentiation method.

## 1. Introduction

Organic scintillators are usually calibrated by means of gamma-lines of radionuclide sources. Due to the low-Z content of the organic scintillators, their response is almost free of any photopeak, constituting a relatively wide Compton edge in the observed spectrum. One major obstacle to this goal, is to accurately determine location of the Compton edge. Several authors have proposed different methods, none of them met the requirements of a powerful calibration procedure; as being accurate, fast and robust. Usually, authors assume a specific location for the Compton edge, in comparison to the value of the count-rate at the location of local maximum, usually described by how much percent it deviates from the local maximum value. For example Beghian et al. [1] suggested 66%, Honecker and Grässler [2] suggested 70%, Bertin et al. [3] suggested 85% to 87%, Knox and Miller [4] suggested 89%, and Swiderski et al. [5] proposed 78% to 82%. Nonetheless this controversial suggestions has a root in the differences between the effective resolution of their detection system used [6]. There are some alternative methods to partially overcome such difficulties. For example, Chikkur and Umakantha [7], based on ad hoc arguments, have proposed to fit a  Gaussian curve to the latter part of the spectrum, putting forward a way to relate the fitting parameters to the Compton edge.

To overcome the abovementioned difficulties, Dietze and Klein [8, 9] have introduced a calibration procedure that is based on simulation of the ideal response by means of a Monte Carlo code; convolving the simulated response into an *assumed Gaussian function*; with the hope that the convolved curve would find appropriate accommodation to the experimental data. Usually, finding the best conformity needs great deal of work, demanding several prediction/correction steps for evaluation of the appropriate Gaussian parameters. Despite its drawbacks, it is more accurate than other empirical methods (discussed earlier) and, more importantly, it could not only determine the Compton edge, but also could define the energy resolution of the detector [6].

Here, based on simple mathematical arguments, a straightforward and accurate method will be presented to determine the Compton edge, as well as the detector resolution. The method, mostly benefited from the fact that an edge in the response could be modeled by a Heaviside step function (HSF), allowing one to find a simple explanation for the derivative of the response. This procedure, is exploited in Section 2. Section 3, demonstrates the applicability of the method by examining some experimental measurements. Section 4 addresses a discussion about the simulation aspects of the Compton scattering, principally directed towards studying the Doppler resolution effects in the Compton scattering process. With the aid of the precision of the *differentiation method*, we are able to quantitatively determine contribution of the Doppler effect into the detector resolution. In this section, we have also addressed a comparison against the measured intrinsic resolution of organic scintillators, reported by Swiderski et al. [10, 11]. Finally, some concluding remarks will be provided in Section 5.

## 2.  Methods

Heaviside step function (HSF) described as follows

$$H(E) = \begin{cases} 1 & E \leq E_c \\ 0 & E > E_c \end{cases}, \tag{1}$$

has the property that its derivative could be described by the Dirac delta function, i.e.,

$$\frac{d}{dE}H(E) = \delta(E - E_c). \tag{2}$$

Interestingly, it could be shown that the derivative of a Gaussian-folded HSF, is also a Gaussian function with similar parameters. Attributing the HSF as a representation of the recoiled proton spectrum, Kornilov et al. [12] have proposed a straightforward method for determination of the maximum recoiled proton energy in an organic scintillator. While the real-world spectrua are far from being a simple HSF, the method is proved to be accurate, because it is based on just a simple operation of differentiation. However, we have noticed that this aspect of the HSF, could be generalized to every other disjoint function. This point will be elaborated below.

The measured spectra are subject to several sources of uncertainties (i.e., errors) which could be depicted by a Gaussian normal distribution. Most of these errors are uncorrelated, making it meaningful to describe the whole set of broadening effects by means of a single Gaussian with an overall estimation for its variance ($\sigma$). A Gaussian normal distribution could be written as

$$G(E) = \frac{1}{\sqrt{2\pi}\sigma} \exp\left[-\frac{1}{2}\frac{E^2}{\sigma^2}\right]. \tag{3}$$

From a mathematical point of view, the *real* response function $R(E)$ emerges from convolution of the *ideal* response function $r(E)$ into the Gaussian distribution,

$$R(E) = r(E) \otimes G(E) = \int_{-\infty}^{+\infty} r(x)G(E-x)dx,$$

The general theme of this work is based on the fact that –frankly speaking– the derivative of every disjoining function (like the HSF), behaves somehow like a Dirac delta or Gaussian functions. One simple and still useful model for an *ideal* Compton edge could be described by the following disjoint second-order function

$$r(E) = \begin{cases} aE^2 + bE + c & E \le E_c \\ 0 & E > E_c \end{cases}, \tag{4}$$

visualized in Figure 1. Its maximum value occurs at $E_c$, to have

$$r_{\max} = aE_c^2 + bE_c + c, \tag{5}$$

meaning that in an ideal system (i.e. not subject to the finite energy resolution effects), the Compton edge is exactly located at the local maximum of the response function. Inclusion of the resolution, demands the convolution of $r(E)$ into a corresponding Gaussian function (3) which results in

$$R(E) = \alpha_1 \cdot \text{erfc}\left[\frac{E-E_c}{\sqrt{2}\sigma}\right] + \beta_1 \cdot \exp\left[-\frac{(E-E_c)^2}{2\sigma^2}\right], \tag{6}$$

where

$$\text{erfc}(E) = 1 - \frac{2}{\sqrt{\pi}} \int_0^E e^{-x^2}dx,$$

is the complementary error function, and the following conventions have been assumed:

$$\alpha_1(E) \equiv \frac{1}{2}\left[a(E^2+\sigma^2) + bE + c\right],$$
$$\beta_1(E) \equiv \frac{-\sigma}{\sqrt{2\pi}}\ a(E+E_c) + b\ . \tag{7}$$

This convolved response (standing for the *real* response function) is of smoother behavior, also displayed in Figure 1. It is simple to verify that the value of $R(E)$ at the Compton edge could be obtained from the following relation:

$$R(E_c) = \alpha_1(E_c) + \beta_1(E_c)$$
$$= \frac{1}{2}r_{max} - \frac{1}{\sqrt{2\pi}} \; 2aE_c + b \; \sigma + \frac{a}{2}\sigma^2 \tag{8}$$

This, strictly shows that the half-value of (or some other percent of) $r_{max}$ could not be considered as a universal measure of the Compton edge ($E_c$). It also indicates that a wider Gaussian resolution (i.e., increasing $\sigma$), shifts the local maximum to the left-hand-side of its original (correct) position, making a sense about the various heuristically reported values for the Compton edge location, described earlier in Section 1.

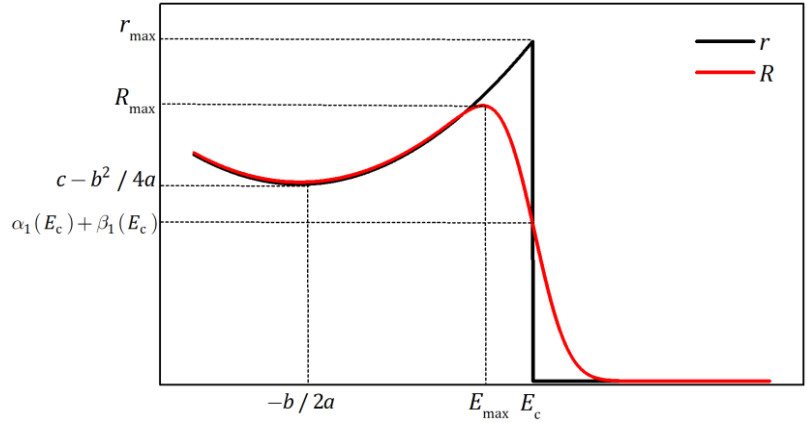

Figure 1. Ideal linear response function: $r(E)$, and the convolved response function: $R(E)$.

Differentiating $R(E)$ we find

$$R'(E) = X_1(E) + X_2(E), \tag{9}$$

where we have accepted the following notation

$$X_1(E) \equiv \alpha_2 \cdot \text{erfc}\left[\frac{E - E_c}{\sqrt{2}\sigma}\right],$$
$$X_2(E) \equiv \beta_2 \cdot \exp\left[-\frac{(E - E_c)^2}{2\sigma^2}\right], \tag{10}$$

and

$$\alpha_2 \equiv \frac{1}{2}(2aE + b),$$
$$\beta_2 = -\frac{1}{\sqrt{2\pi}\sigma}\left[a(E_c^2 + 2\sigma^2) + bE_c + c\right]. \tag{11}$$

The shape of these latter functions are illustrated in Figure 2, and they pose interesting relations. Especially there is a constant slope at the rightmost, governed by $X_1$ as follows

$$\theta = a \cdot \operatorname{erfc}\left[-\frac{E_c}{\sqrt{2}\sigma}\right] \tag{12}$$

Although the shape of $R'$ is mostly governed by $X_2$ (which indeed is a Gaussian distribution), one can conclude from Figure 2 that the correct location of local minimum slightly differs from that of the $X_2$. The correct value of this minimum could be obtained by

$$E_{\min} \simeq E_c + \frac{b\sigma^2 - \sqrt{2\pi}a\sigma^3}{a(E_c^2 - \sigma^2) + bE_c + c} \tag{13}$$

This necessitates a positive correction term (usually below 1%), as it also could be seen in Figure 2. Moreover, this equation indicates that increasing the Gaussian variance (i.e., deteriorating the resolution), as well as decreasing the incident energy, magnifies this correction term. Although, it is useful to approximate the Compton edge, solely by the local minimum of the $X_2$

$$E_c \simeq E_{\min} \tag{14}$$

This approximation is acceptable for most interesting cases, because usually we have $\sigma \sim 1\,\text{keV}$, while $E_c \sim 10 - 100\,\text{keV}$, diminishing the correction term. Noting this, one is able to calculate the approximate $R'_{\min} \approx R'(E_c)$ as

$$R'_{\min} = aE_c + \frac{b}{2} - \frac{1}{\sqrt{2\pi}\sigma}\left[a(E_c^2 + \sigma^2) + bE_c + c\right]$$

## 3. Experiments

For experimental verification of the differentiation method, we have measured the response function of a ∅2"×2" BC400 plastic scintillator [13] irradiated by three gamma-ray sources: [137]Cs, [22]Na and [60]Co (see Table 1). Due to the lack of enough resolution (in our detection system) to discriminate between distinct [60]Co gamma-lines, we assumed an average energy of 1252.8 keV for this source.

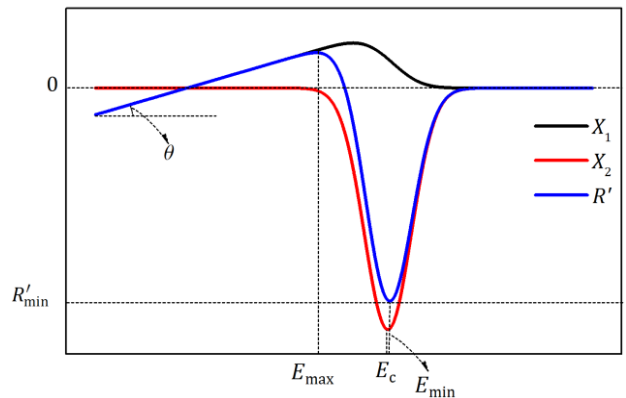

Figure 2. Differentiation of the ideal and the real response functions

Table 1. Gamma-ray sources for experiments.

| Nuclide | ¹³⁷Cs | ²²Na | ⁶⁰Co |
|---|---|---|---|
| **Energy (keV)** | 661.7 | 511.0 | 1173.2 |
| | | 1274.5 | 1332.5 |
| | | | Average: 1252.8 |
| **Exact Compton edge (keV)** | 477.0 | 339.0 | 963.5 |
| | | 1061.0 | 1332.5 |
| | | | Average: 1040.2 |
| **Estimated Compton edge (keV)** | 477.2±0.06 | 339.5±0.07 | - |
| | | 1061.0±0.05 | - |
| | | | Average: 1042.2±0.06 |
| **Correct location (%)** | 57.4 | 56.3 | - |
| | | 68.0 | - |
| | | | 65.1 |
| **Resolution (%)** | 14.06±0.01 | 15.96±0.01 | - |
| | | 11.97±0.02 | - |
| | | | 11.87±0.03 |

Undoubtedly, it is expectable to find the measured (or even simulated spectra) that show unavoidable (but might be invisible) fluctuations, resulting in serious fluctuations in their corresponding derivatives. Shown in Figure 3 is the measured spectrum of ¹³⁷Cs, along with its corresponding differentiation. However, it is notable that most of these fluctuations are of non-correlated nature and could be successfully eliminated by *low-pass filtering* with appropriate windowing (here Blackman window), keeping the original progression of the data [14]. The cut-off angular frequency ($\omega_c$) of the low-pass filter could vary over the domain $[0, \pi]$, implying variation of the output, ranging from the no-pass (for zero) to all-pass (for $\pi$) filters. With the aid of a robust nonlinear fitting procedure, there would be no troubles in fitting, even with a noisy data. Figure 4 demonstrates this point, because it shows that low-pass filter smoothing has no considerable effect on the outcome of the procedure (mostly about ±1%). This is somehow contrary to Kornilov et al. [12] and Stevanato et al. [15], whom found problems with raw (non-filtered) data. For the fitting purposes one could focus on the central region around the appeared peak, which is useful to discard asymmetry effects of the data, shown by shaded region in Figure 3(b) in the case of ¹³⁷Cs. The accuracy of the method could be concluded from the results of Table 1, which explicitly shows that assuming a specific percentage of the local Compton maximum is not a correct representative of the Compton edge location. For example, the correct location is taking 80% of the maximum (as it is usually the case), one gets about 27 keV deviation from the correct location in for $^{137}$Cs peak.

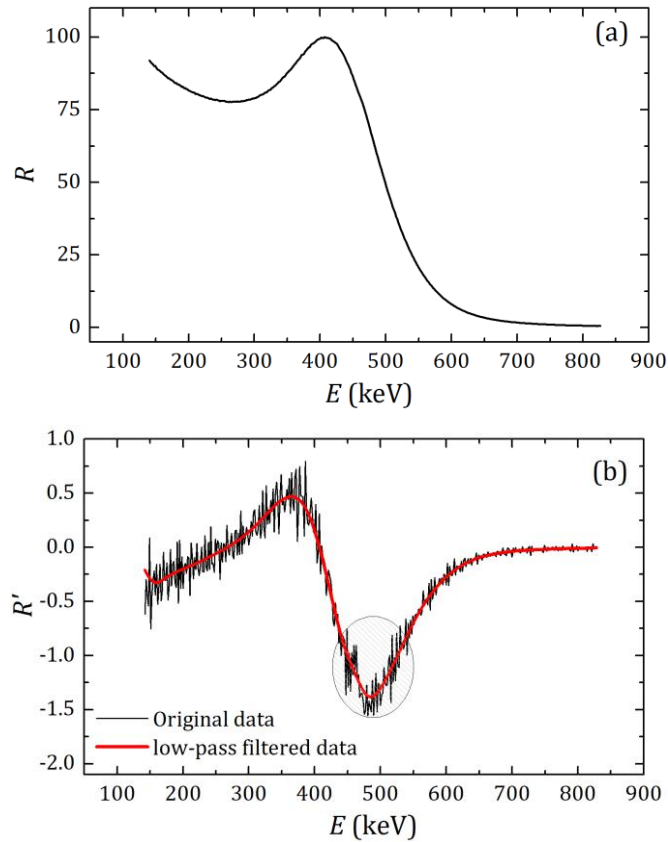

Figure 3. (a) Measured response function for $^{137}$Cs gamma-ray source. (b) Derivative of the measured response function (thin black line), and low-pass filtered response function (thick red line).

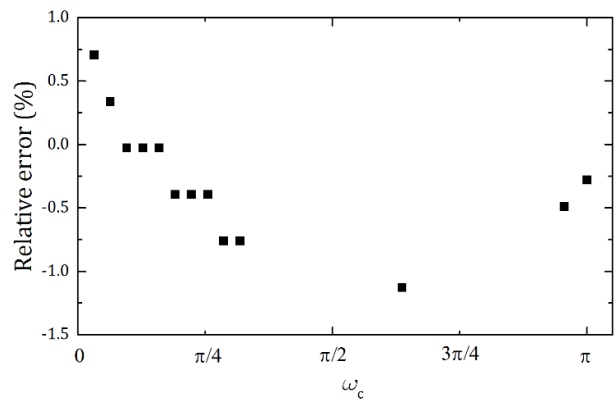

Figure 4. Relative error ($100 \times (E_x - E_c) / E_c$) in the determination Compton edge ($E_x$), after low-pass filtering with different cut-off frequencies ($\omega_c$).

After adjusting the calibration coefficient by means of the $^{22}$Na peaks, we were able to predict the location of the $^{137}$Cs and $^{60}$Co gamma-lines, depicted in Figure 5, with which that uncertainty of the experimental data12 are practically invisible. Moreover, it could be seen that the approximate location of the effective gamma-rays of $^{60}$Co are also acceptable within experimental errors. For comparison purposes, we made a comparison against some other experimental data, reported by Jolivette and Rouze in Ref. [16]. These data were measured by an HPGe detector, so they are of good quality (usually having about 1 keV uncertainties), as it could be seen in Figure 5 by comparing against MCNPX simulations.

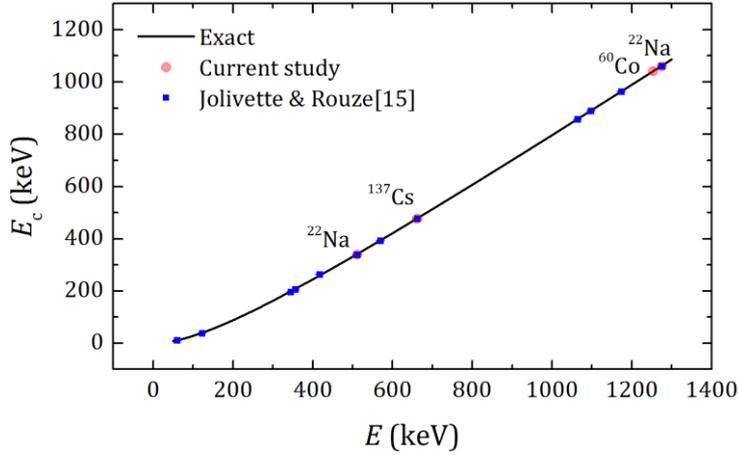

Figure 5. Compton edge location, as specified by the differentiation method (Current study), in comparison to the MCNPX (Exact) estimations and the measurements by Jolivette & Rouze [16].

## 4. Doppler resolution

Bounded atomic electrons are not at rest, making certain deviations from the free-electron assumption in the derivation of the conventional Klein-Nishina cross sections. This boundedness, results in Gaussian broadening of the energy-angle distribution of the outgoing entities, usually referred to as the *Doppler broadening* [17]. It is well-known and well-documented that this Compton scattering, usually implies a finite *Doppler resolution* in the energy distribution of the detector's response function [18]. Notwithstanding, this Doppler resolution completely differs from other contributing effects in the total energy resolution of a scintillation detector (such as the light generation and transport effects, electronic noise/fluctuations effects, etc.), revealing a situation somehow similar to the *intrinsic resolution* in the case of inorganic scintillators. While this property basically could be asserted, up to our knowledge, there is no quantitative estimation for it, which is deemed to be due to the lack of a precise enough method for accurate localization of the Compton edge. Here in this section, we are aiming to resolve this issue by employing the advantages of the *differentiation method*.

Most of the modern radiation transport simulation codes have certain capabilities to handle low-energy photon interactions perceiving the momentum distribution of bounded electrons. For example, there were immense efforts to develop such a capability in GEANT4 [19, 20], FLUKA [21, 22], EGS4 [23-25], PENELOPE [26-29], and especially in MCNPX [30, 31].

MCNPX/5 is a multi-purpose Monte Carlo simulation code [31] with the capability to handle bounded electrons [32-34]. Here in this study, we were made use of the MCNPX code to simulate the response function, taking note of the shortcomings of its previously released photons cross section library [35, 36], hence considering its more recent version [37].

Considering an ideal ∅2"×2" BC400 plastic scintillator, we have performed the simulations reported below. According to the BC400 datasheet [13] the material composition were fixed to H/C:1.103, mass density: 1.032.

Before consulting to the detector simulation, we have checked the simulation of bounded electrons for 500 keV (Figure 6) and 30 keV (Figure 7) gamma-rays, by considering the Compton interaction in a tiny sample (~$10^{-5}$ cm) of the plastic scintillator. As a matter of study, we have calculated the intensity of the outgoing photons as a double-differential function of energy and angle. These figures not only help to explicitly visualize the broadening effects of the bounded electrons, but also permitting us to visually observe that the broadening is not uniform for all outgoing angles (see Figure 7b).

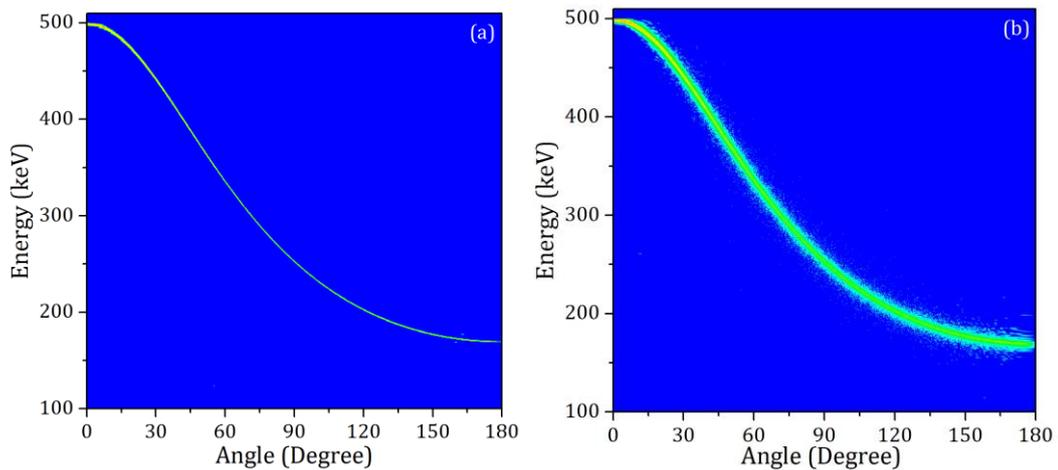

Figure 6. Double-differential distribution of the outgoing flux of gamma-rays as a function energy and angle, for incident photon $E$=500 keV, (a) without and, (b) with the bounded electron model.

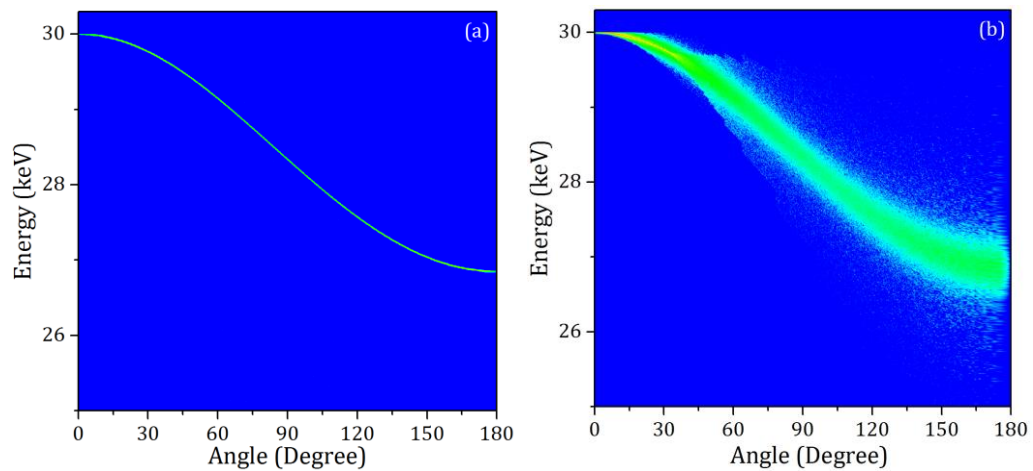

Figure 7. Same as Figure 6, for incident photons with $E$=30 keV.

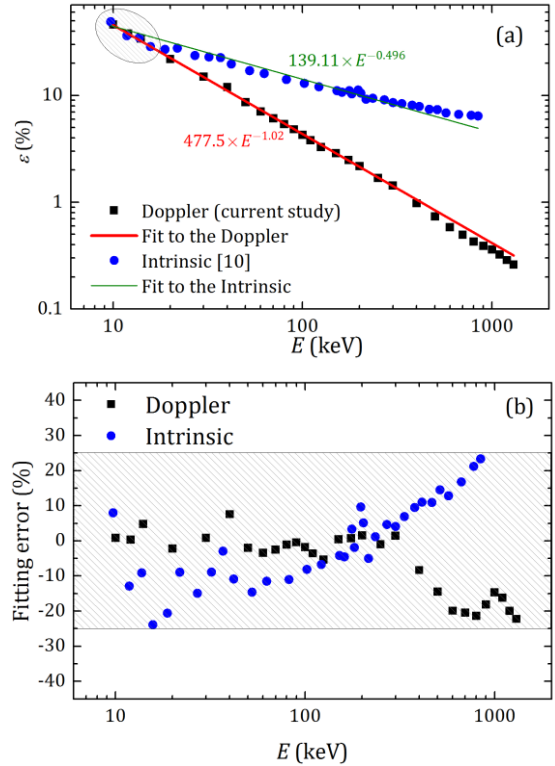

Figure 8. (a) Energy resolution due to the Doppler broadening. The continuous line represents the fitted curve. (b) Corresponding error in the fitted curves.

By applying the *differentiation method* for analysis of the simulated spectra we have determined the FWHM and the corresponding energy resolution (Figure 8) that are exclusively resulted from the bounded electron effects. The relation between resolution ($\varepsilon$), the FWHM and the variance of the Gaussian peak ($\sigma$) is the following well-known equation [18]

$$\varepsilon(\%) = 100 \times \frac{\text{FWHM}}{E_0} = 200\sqrt{2\ln 2}\,\frac{\sigma}{E_0}\ . \tag{15}$$

Fitting the Doppler resolution using a power model function of the form

$$\varepsilon = \alpha E^{\beta}, \tag{16}$$

we find $\alpha = 477.5 \pm 0.03$ and $\beta = -1.02 \pm 0.02$, persuading the $1/E$ relation for the Doppler resolution. It is noteworthy that the error[1] in the fitted curve is mostly below 25%, while it is more accurate at lower energies (below ∼300 keV), depicted in the dashed region of Figure 8(b).

The estimated Doppler resolution varies from 1.5% to 45% for incident photons of energies below about 300 keV down to 10 keV. This has a strict implication, urging to incorporate the Doppler broadening -or detailed physics model in terms of the MCNPX' nomenclature- into the simulations. One should not forget that the detailed physics model of the MCNPX code is *disabled* by default. Swiderski et al. have studied the intrinsic resolution of organic scintillators [10], taking all possible effects into account. Their measurements for BC408 resulted in a set of data which are consistent with the fitting of power model Eq. (16), with $\alpha = 139.11 \pm 0.02$ and $\beta = -0.496 \pm 01$ . This result is very close to the $1/\sqrt{E}$ relation, as it could be anticipated by noting the well-known role of the statistical uncertainty in the energy resolution which is related to the population of received optical photons. Moreover, both data tend to similar values at lower energies ($\sim 10$ keV) that permits one to conclude about the significant role of the Doppler effect in such situations. It could be observed that the $1/E$ nature is more profound below 20 keV, emphasized by shaded region in Figure 8(a), which is found to be well-fitted by Eq. (16) with these parameters: $\alpha = 418.82 \pm 0.01$ and $\beta = -0.96 \pm 0.02$ .

---

[1] Relative error in every fitted curve has been calculated with the following equation: $100 \times (\varepsilon - \varepsilon_{\text{Fit}}) / \varepsilon$.

## 5. Conclusions

A simple and accurate method based on the differentiation of the response function was proposed and detailed for localization of the Compton edge, which also is able to formalize the energy resolution of the detector. The method examined and proved to be accurate in dealing with the experimental measurements, viz., analyzing the response of a BC400 detector irradiated by three different gamma-ray sources. Based on the extensive Monte Carlo simulations, we have determined the intrinsic energy resolution of an organic scintillator, which was shown to be principally originated from the Doppler effects.